\documentclass[aps,prb,preprint,longbibliography,superscriptaddress]{revtex4-2}
\usepackage{mathrsfs}
\usepackage{amsmath,gensymb}
\usepackage{amsfonts}
\usepackage{amssymb}
\usepackage{amsthm}
\usepackage{graphicx}
\usepackage{natbib}
\usepackage{xcolor}
\usepackage{hyperref}
\usepackage{bm}
\usepackage[caption=false]{subfig}
\usepackage{verbatim}
\usepackage[utf8]{inputenc}
\usepackage{soul}

\newcommand{\nocomments}{\long\def\comm##1\commend{}}

\nocomments

\newcommand{\akcom}[1]{{\comm \color{red} Akash: ``#1'' \commend}}

\newcommand{\simcom}[1]{\comm \textcolor{green}{Simran: ``{#1}''} \commend}

\newcommand{\lina}[1]{{\color{black} #1}}
\newcommand{\ak}[1]{{\color{black} #1}}
\newcommand{\add}[1]{{\color{black} #1}}

\begin{document}
\title{Generation of spin-triplet Cooper pairs via a canted antiferromagnet}

\author{Simran Chourasia}
\email{simran.chourasia@uam.es}
\affiliation{Condensed Matter Physics Center (IFIMAC) and Departamento de F\'{i}sica Te\'{o}rica de la Materia Condensada, Universidad Aut\'{o}noma de Madrid, E-28049 Madrid, Spain}

\author{Lina Johnsen Kamra}
\affiliation{Center for Quantum Spintronics, Department of Physics, Norwegian University of Science and Technology, NO-7491 Trondheim, Norway}

\author{Irina V. Bobkova}
\affiliation{Moscow Institute of Physics and Technology, Dolgoprudny, 141700 Moscow, Russia}
\affiliation{National Research University Higher School of Economics, Moscow, 101000 Russia}

\author{Akashdeep Kamra}
\affiliation{Condensed Matter Physics Center (IFIMAC) and Departamento de F\'{i}sica Te\'{o}rica de la Materia Condensada, Universidad Aut\'{o}noma de Madrid, E-28049 Madrid, Spain}

\begin{abstract}
Spinful triplet Cooper pairs can be generated from their singlet counterparts available in a conventional superconductor (S) using two or more noncollinear magnetic moments, typically contributed by different magnets in a multilayered heterostructure. Here, we theoretically demonstrate that an S interfaced with a canted antiferromagnet (AF) harbors spinful triplet Cooper pairs capitalizing on the intrinsic noncollinearity between the two AF sublattice magnetizations. As the AF canting can be controlled by an applied field, our work proposes a simple bilayer structure that admits controllable generation of spin-triplet Cooper pairs. Employing the Bogoliubov-de Gennes framework, we delineate the spatial dependence of the spin-triplet correlations. We further evaluate the superconducting critical temperature as a function of the AF canting, which provides one experimental observable associated with the emergence of these triplet correlations. 
\end{abstract}

\maketitle

\section{Introduction}\label{sec:intro}

The dissipationless flow of charge in superconductors is partly responsible for their central role in the various emerging quantum technologies~\cite{Xiang2013,Krantz2019}. The widely available conventional superconductors, such as Al and Nb, are made of spin-singlet Cooper pairs, which harbor no net spin~\cite{Sigrist1991}. A superconductor hosting spin-triplet Cooper pairs can support dissipationless spin currents~\cite{Eschrig2015,Bergeret2005,Bergeret2018,Buzdin2005,Linder2015}, deemed valuable for switching magnetic memories~\cite{Guang2021,Bobkova2018,Waintal2002,Halterman2016,Linder2011}, as well as exotic excitations, such as Majorana bound states~\cite{Fu2008,Tanaka2012}\akcom{\cite{Tanaka2012} has been added.}. Such an unconventional superconductor can be engineered from its conventional counterpart employing heterostructures incorporating magnetic multilayers~\cite{Bergeret2005,Eschrig2008,Buzdin2005,Linder2015,Bergeret2018}. The basic requirement for achieving spinful triplets from spin-singlets is exposing the latter to two or more noncollinear spin-splitting fields. A wide variety of multilayered hybrids comprising conventional superconductors (S) and ferromagnets (F) has been employed to achieve the desired spinful triplets~\cite{Keizer2006,Khaire2010,Jeon2018,Jeon2020,Diesch2018,Costa2022,Hijano2022,Robinson2010,Chiodi2013,Ruano2020, Wu2012b, Banerjee2018, Johnsen2019, Jacobsen2015, Tamura2023,Dutta2019,Leksin2012}\simcom{(here, \cite{Tamura2023,Dutta2019} have been added)}, coming a long way from the initial critical temperature studies~\cite{Buzdin1990,Radovic1991}.

Since a N\'eel ordered antiferromagnet (AF) bears no net spin or magnetic moment, for some time it was considered inert at causing spin-splitting in an adjacent superconductor. Indeed, early experiments found a metallic AF to behave just like a normal metal when considering its effect on an adjacent superconductor~\cite{Werthammer1966}. More recent experiments, on the other hand, found the AF to substantially affect the adjacent S with intriguing dependencies~\cite{Hubener2002,Bell2003,Wu2013,Seeger2021,Mani2009,Mani2015}. From the theory perspective, Josephson junctions~\cite{Bobkova2005,Andersen2006} and interfaces~\cite{Andersen2005,Jakobsen2020}\simcom{(here,\cite{Jakobsen2020} is added)} involving itinerant AFs were shown to exhibit non-trivial properties due to quasiparticle reflections. Moreover, a recent work demonstrates that an uncompensated interface of an insulating AF with an adjacent S induces a strong spin-splitting as well as spin-flip scattering thereby strongly influencing the S~\cite{Kamra2018}. Subsequent work found even the fully compensated interface between the AF and S to be spin-active~\cite{Johnsen2021}. This has been understood as due to the AF inducing N\'eel triplets whose pairing amplitude has an alternating sign in space similar to the AF spin~\cite{Bobkov2022}. Altogether, the potential usefulness of AFs~\cite{Baltz2016} in engineering novel superconducting effects and devices, such as a filter~\cite{Fyhn_arxiv}, is starting to be understood.    

Pekar and Rashba~\cite{Pekar1964} already recognized long ago that even though the net spin vanishes in an AF, at the lattice constant length scale, the AF harbors a spin or magnetization profile that rapidly varies in space changing its sign from one lattice site to the next, which should manifest itself in physical observables. As a result, the AF with its two sublattice magnetizations antiparallel to each other generates zero-spin N\'eel triplets~\cite{Bobkov2022}. Proceeding further along this line of thought, a homogeneous canted AF with its sublattice magnetizations deviating from an antiparallel alignment effectively harbors a noncollinear spin texture capable of generating spinful triplet Cooper pairs in an adjacent S. This exciting possibility is theoretically examined in the present work. \ak{Furthermore, a recent experiment~\cite{Jeon2023} demonstrates generation and use of spinful triplet correlations employing the intrinsically noncollinear ground state of a kagome AF.}

Here, employing the Bogoliubov-de Gennes (BdG) framework~\cite{Zhu2016}, we theoretically investigate a bilayer structure consisting of an insulating AF exchange coupled via a compensated interface to an adjacent S. We examine the critical temperature and spin-triplet correlations in the S as a function of the canting in the AF, which allows us to continuously tune the AF from its collinear antiparallel state to it effectively becoming an F. We find that N\'eel triplets are generated in the S both from the interband pairing channel considered recently~\cite{Bobkov2022}, and from the conventional intraband pairing. \ak{The former channel is dominant at half-filling in which case states from two different electronic bands are energetically close to the chemical potential and can participate in forming these unconventional N\'eel triplet Cooper pairs~\cite{Bobkov2022}. This can be compared with pairing at finite energies in other multisublattice systems~\cite{Tang2021}. The conventional intraband pairing channel dominates away from half-filling when only states within the same electronic band are energetically close to each other and the chemical potential.} We find that the \ak{intraband pairing} channel results in N\'eel triplets formation due to an imprinting of the N\'eel character by the AF on the normal state electronic wavefunctions in the S. Although this channel of N\'eel triplet generation is found to be much weaker than the interband pairing channel, it admits qualitatively new effects. We show that it is only in the intraband pairing channel that spinful N\'eel triplets are generated due to the intrinsic noncollinearity of a canted AF. The S critical temperature variation as a function of the AF canting angle is found to be consistent with the intraband N\'eel triplets being weaker than their interband counterparts, and may offer a convenient experimental signature of this interplay.

The paper is organized as follows. Section \ref{model} introduces the model and BdG framework employed in our analysis. The dependence of spin-triplet correlations on space and AF canting is discussed in Sec.~\ref{Tripcorr}, while the variation of superconducting critical temperature is discussed in Sec.~\ref{sec:Tc}. Until this point, we consider a one-dimensional model that allows a simple and semianalytic understanding of the essential physics. In Sec.~\ref{sec:2D}, we employ a two-dimensional \ak{(2-D)} model for the S to validate our prior results and further examine the spatial dependence of the \ak{s-wave and p-wave} triplet correlations. We conclude with discussion and summary of the key points in Sec.~\ref{sec:summary}. The appendices provide details of the BdG framework, analytic evaluation of the normal state electronic properties in the bilayer, \lina{a discussion of the decay length of the spin-triplet correlations inside the S, \ak{critical temperature} results for the 2-D system,} triplet correlations for a different configuration of the AF sublattice magnetizations, and the parameters employed in our numerical routines.


\section{AF/S bilayer model} \label{model}

\begin{figure}[tb]
	\begin{center} 
		\includegraphics[width=85mm]{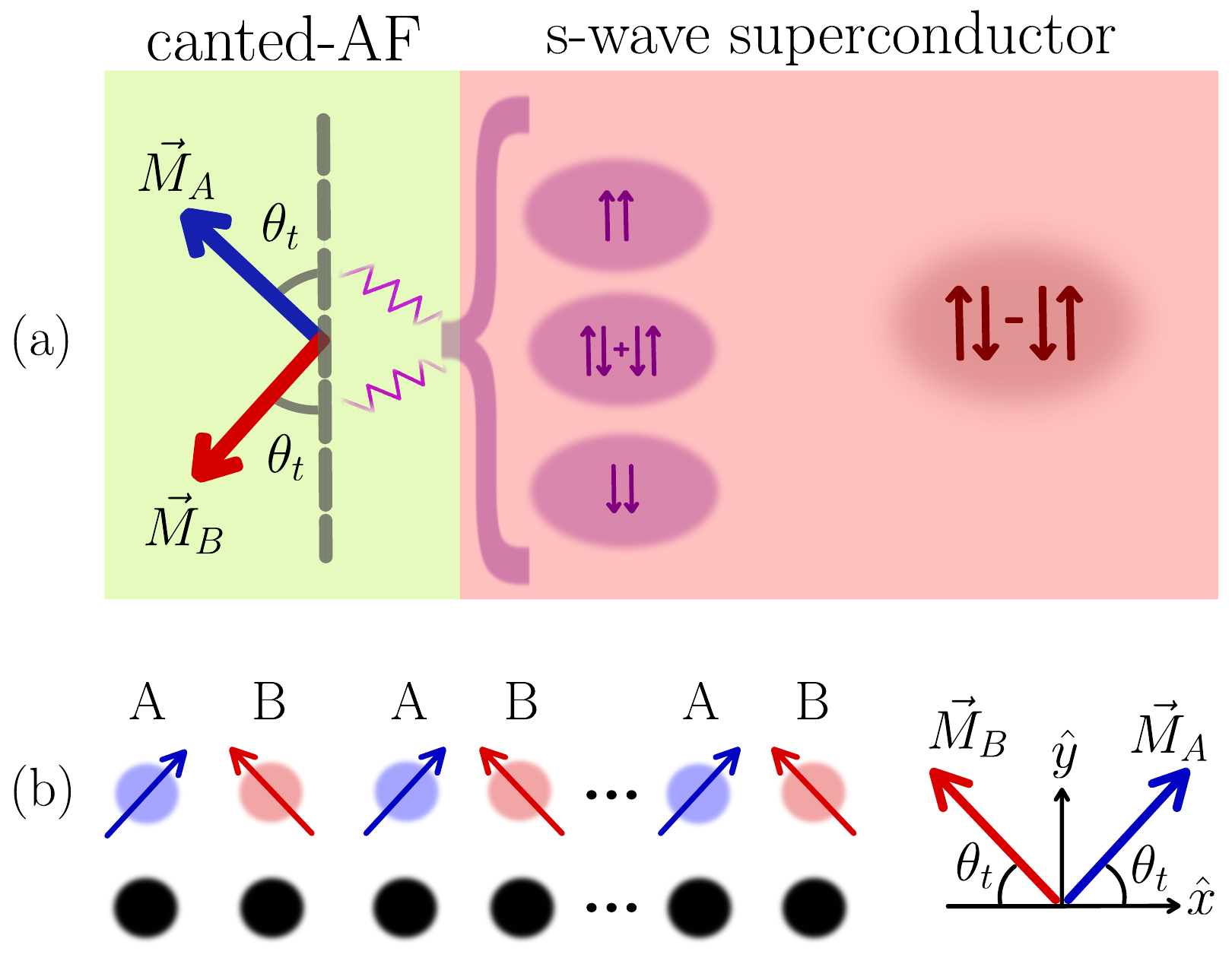}
		\caption{(a) Schematic depiction of the system and key physics under investigation. Equal-spin and zero-spin triplet correlations are generated in a conventional $s$-wave spin-singlet superconductor when it is interfaced with a canted antiferromagnet (canted-AF). This results from the intrinsic noncollinearity between the two AF sublattice magnetizations. (b) Schematic depiction of our model investigated using the Bogoliubov-de Gennes method. The black circles represent lattice sites of the superconductor \ak{hosting itinerant electrons and Cooper pairs. Blue} (red) circles represent A (B) sublattice sites of the \ak{electrically insulating} canted-AF. The blue and red arrows denote the local AF magnetic moments. The canting angle $\theta_t$ allows us to vary the magnet from being a collinear AF ($\theta_t = 0$) to a ferromagnet ($\theta_t = \pi/2$).}
		\label{fig:1}
	\end{center}
\end{figure}

We consider a bilayer structure comprising an insulating AF exchange coupled via a compensated interface to the S layer, as depicted schematically in Fig.~\ref{fig:1}(a). We anticipate that all three kinds of spin-triplet correlations will be generated in the S when the AF sublattice magnetizations are canted [Fig.~\ref{fig:1}(a)]. We employ the Bogoliubov-de Gennes method and numerically evaluate the superconducting properties self-consistently~\cite{Zhu2016}. The canted-AF is taken to be an ideal insulator with a large band-gap. Consequently, the Hamiltonian is formulated only for the itinerant electrons in S \ak{as they never enter the insulating AF}. The AF's influence on the S is accounted for by incorporating a spatially dependent spin-splitting caused by the AF spins~\cite{Kamra2018,Cottet2009,Eschrig2015b}. Furthermore, with the aim of allowing a semianalytic understanding to the essential physics, we consider a one-dimensional model as depicted schematically in Fig.~\ref{fig:1}(b). The resulting Hamiltonian is given by
\begin{align} \label{eq1}
 H = &  -\mu \sum_{j,\sigma} c_{j,\sigma}^\dagger c_{j,\sigma} -t \sum_{\left<i,j\right>}\sum_{\sigma} c_{i,\sigma}^\dagger c_{j,\sigma}-\frac{J}{2} \sum_{j} \vec{M}_j \cdot \vec{S}_j  \nonumber\\ 
 & +\sum_j \left( \frac{|\Delta_j|^2}{U} + \Delta_j^* c_{j,\downarrow} c_{j,\uparrow} + \Delta_j c_{j,\uparrow}^\dagger c_{j,\downarrow}^\dagger \right),
\end{align}
where $c_{j,\sigma}^\dagger$ ($c_{j,\sigma}$) is the creation (annihilation) operator of an electron with spin $\sigma$ at site $j$ of the S layer, with $j=1,2,\dots,N$ as the site index. Here, the spin quantization axis is taken along the $z$-axis. We further consider periodic boundary conditions by allowing electrons to hop between sites $j=1$ and $j=N$.

In Eq.~\eqref{eq1}, $\mu$ is the chemical potential. Within our theoretical method, it is determined via the filling factor $f$, which is the fraction of filled electronic states in the system. For $f=0.5$ (half-filled band), we obtain $\mu=0$. This corresponds to the Fermi wave-vector $k_F = \pi/2a$ located at the AF Brillouin zone boundary, where $a$ is the lattice constant. The normal state electronic dispersion and properties for the case of an AF in its collinear anti-parallel state have been discussed in Appendix~\ref{AFNsection}. For $f \neq 0.5$, $\mu$ is non-zero and the Fermi level is away from the AF Brillouin zone boundary. In the following analysis, we will consider the two qualitatively distinct cases of $f = 0.5$ and $f \neq 0.5$, corresponding to $\mu = 0$ and $\mu \neq 0$.

The second term in Eq.~\eqref{eq1} is the kinetic energy term, describing hopping between nearest neighboring sites $\left<i,j\right>$ with $t>0$ as the hopping parameter.

The third term in the Hamiltonian [Eq.~\eqref{eq1}] accounts for the spin-splitting due to the localized magnetic moments in the canted-AF~\cite{Kiwi2001,Bergeret2018,Maki1964,Kamra2018}. $J\vec{M}_j/2$ is the local spin-splitting field which causes an energy shift of spin-up and spin-down electrons by $\mp J/2$, with respect to the local spin-quantization axis along $\vec{M}_j$. \ak{The strength of this spin-splitting, and thus the value of $J$ in our model, depends on the S thickness and several other parameters~\cite{Kamra2018}. It can thus be tuned in a broad range via appropriate thin films fabrication~\cite{Kamra2018} as per the experimental requirements.} Here, $\vec{M}_j=[(-1)^{j+1}\cos \theta_t \, \hat{x} + \sin \theta_t \, \hat{y}]$ is the unit vector along the direction of local magnetic moment at the $j^{th}$ canted-AF site. The AF sublattice magnetizations are taken to be in the $x$-$y$ plane to examine the spin-triplet correlation with the quantization axis ($z$) perpendicular to this plane. A different magnetic configuration has also been investigated and discussed in the appendix. $\vec{S}_j=[(c_{j,\uparrow}^\dagger c_{j,\uparrow}-c_{j,\downarrow}^\dagger c_{j, \downarrow})\hat{z}+(c_{j,\uparrow}^\dagger c_{j,\downarrow}+c_{j,\downarrow}^\dagger c_{j,\uparrow}) \hat{x} + (-i c_{j,\uparrow}^\dagger c_{j,\downarrow}+i c_{j,\downarrow}^\dagger c_{j,\uparrow})\hat{y}]$ is the spin operator of an electron at site $j$ of the superconductor. 

The last term in Eq.~\eqref{eq1} accounts for the conventional $s$-wave spin-singlet superconducting correlations. It is obtained by mean field approximation of the pairing interaction $-U \sum_j n_{j,\uparrow} n_{j,\downarrow}$, where $U>0$ is the attractive pairing potential and $n_{j,\sigma}=c_{j,\sigma}^\dagger c_{j,\sigma}$ is the number operator~\cite{Zhu2016}. $\Delta_j=-U\langle c_{j,\downarrow} c_{j,\uparrow}\rangle$ is the resulting superconducting order parameter.

The total Hamiltonian Eq.~\eqref{eq1} is numerically diagonalized and the superconducting state is determined self-consistently as detailed in Appendix~\ref{AC}. The exact parameters employed in our numerical routines have been specified in Appendix~\ref{parameter}.


\section{Triplet correlations} \label{Tripcorr}

In this section, we quantify and investigate the different spin-triplet correlations in the S. Consider the anomalous Matsubara Green's function $F_{jj,\sigma\sigma'}(\tau)=-\langle T_\tau c_{j,\sigma}(\tau) c_{j,\sigma'}(0) \rangle$, where $\tau=i\tilde{t}$ is the imaginary time with $\tilde{t}$ as the time~\cite{Kopnin2001,Belzig1999}. Further, $T_{\tau}$ is the ordering operator for imaginary time $\tau$. In the Fourier space, we obtain
\begin{align}
F_{jj,\sigma\sigma'}(i\omega_l)=\int_0^{\beta} e^{i\omega_l \tau} F_{jj,\sigma\sigma'}(\tau) ~  d\tau,
\end{align}
where $\beta=\hbar/k_B T$, $k_B$ is the Boltzmann constant, $T$ is the temperature, and $\omega_l=(2l+1)\pi/\beta$ are the fermionic Matsubara frequencies with integer $l$. See Appendix~\ref{AC} for further calculation details. Since the spin-triplet correlations are odd in frequency~\cite{Bergeret2005}, we take a sum over all positive Matsubara frequencies to define an appropriate dimensionless quantity that would allow us to quantify the correlations
\begin{align}\label{corr1}
F_{j,\sigma\sigma'}= \frac{1}{\beta} \sum_{\omega_l>0} F_{jj,\sigma\sigma'}(i\omega_l).
\end{align}
Employing this notation, we express the relevant superconducting correlations
\begin{align}
F^s_j & =\frac{1}{2}\left( F_{j,\downarrow\uparrow}-F_{j,\uparrow\downarrow}\right) \label{corr2},\\
F^{t,z}_j & = -\frac{1}{2}\left( F_{j,\downarrow\uparrow}+F_{j,\uparrow\downarrow} \right) \label{corr3},\\
F^{t,x}_j & = \frac{1}{2} \left(F_{j,\uparrow\uparrow}-F_{j,\downarrow\downarrow}\right) \label{corr4}, \\
F^{t,y}_j & = \frac{i}{2}\left( F_{j,\uparrow\uparrow}+F_{j,\downarrow\downarrow} \right) \label{corr5}.
\end{align}
 $F^s$ is the spin-singlet correlation. $F^{t,z}$, $F^{t,x}$, and $F^{t,y}$ are the zero-spin triplet correlations when spin is measured along $z$-, $x$-, and $y$-axis respectively. Together, the latter three [Eqs.~\eqref{corr3}-\eqref{corr5}] allow us to express all three kinds of the spin-triplet correlations with $z$ quantization axis. We evaluate the quantities defined in Eqs.~\eqref{corr2}-\eqref{corr5} to investigate the different superconducting correlations in our system.

\subsection{Numerical results}

In an isolated conventional superconductor, only $F^s$ is non-zero while $F^{t,x}$, $F^{t,y}$, and $F^{t,z}$ are zero. Now, if we consider an F/S bilayer, \ak{then the propagating electronic wavefunctions in the normal state of the S film acquire a spin-dependent phase associated with their spin-dependent energies resulting from the spin-splitting induced by the F~\cite{Eschrig2007}. The result is a relative phase difference between the opposite spin electrons as they propagate. In the superconducting state of the S layer, this} causes the zero-spin triplet correlation with spin-quantization axis along the F magnetization to become non-zero~\cite{Eschrig2007,Bergeret2005}. For example, if the magnetization of the ferromagnet is along the $z$-axis, then $F^{t,z}$ becomes non-zero while $F^{t,x}$ and $F^{t,y}$ remain zero. Similarly, $F^{t,x}$ and $F^{t,y}$ become non-zero when magnetization of the ferromagnet is along $x$- and $y$-axis, respectively. 

Quasiclassical theory~\cite{Kopnin2001,Belzig1999} shows that the triplet vector $\vec{F^t_j}=F^{t,x}_j \hat{x} + F^{t,y}_j \hat{y} + F^{t,z}_j \hat{z}$ always has a component aligning with the local exchange field $J\vec{M}_j/2$ whether the magnetization of the ferromagnet in a F/S bilayer is homogeneous or inhomogeneous~\cite{Champel2005, Eschrig2015}. We, therefore, study these correlations in order to decompose the contribution of antiferromagnetic and ferromagnetic components of the canted-AF.

\begin{figure*}[tb]
	\begin{center} 
		\includegraphics[width=\linewidth]{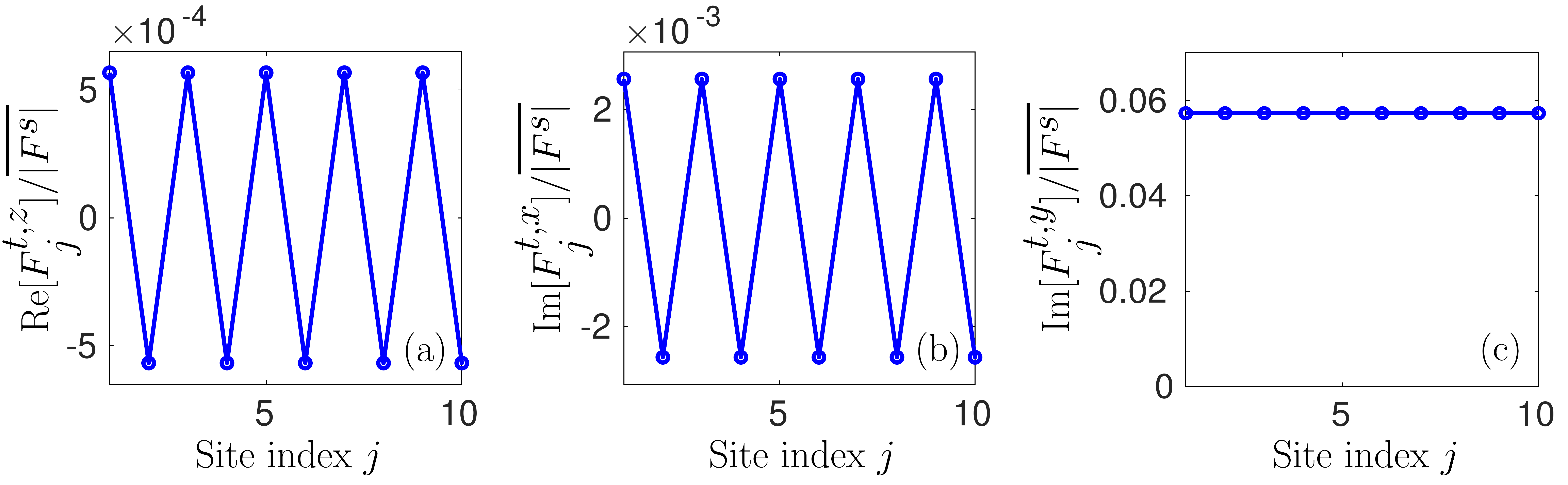}
		\caption{Spatial variation of the normalized triplet correlations for 10 lattice sites considering maximal noncollinearity corresponding to the canting angle $\theta_t = \pi/4$ and filling factor $f=0.6$ ($\mu/\Delta_0 \approx 37$, where $2 \Delta_0$ is the zero-temperature superconducting gap without the adjacent AF). We plot the real part of the zero-spin triplet correlation $F^{t,z}_j$ [panel (a)], and the imaginary parts of the spin-triplet correlations $F^{t,x}_j$ [panel (b)] and $F^{t,y}_j$ [panel (c)]. The imaginary part of the former and the real parts of the latter two are zero. All the correlations are normalized by the spatially averaged magnitude of the singlet correlation $\overline{ |F^s| }$. The detailed parameters employed for the numerical evaluation are specified in Appendix~\ref{parameter}.}
		\label{fig:2}
	\end{center}
\end{figure*}

For $\theta_t=0$ (Fig.~\ref{fig:1}), our considered AF becomes a collinear antiferromagnet with the axis of magnetic moments along the $x$-direction. As we increase the value of $\theta_t$ by a small amount, the canted-AF acquires a net magnetization along the $y$-direction. So the canted-AF can be decomposed into an antiferromagnetic component (along the $x$-axis) and a ferromagnetic component (along the $y$-axis). For a collinear AF, we obtain N\'{e}el triplets. This means the component of $\vec{F^t_j}$ parallel to the axis of the N\'{e}el vector modulates with the N\'{e}el order of the AF~\cite{Bobkov2022}. For $\theta_t = \pi/2$, we effectively obtain a F/S structure with the F magnetization along the $y$-axis.

We now investigate the case of $\theta_t = \pi/4$ that produces maximum noncollinearity between the two AF sublattice magnetization. In Fig.~\ref{fig:2}, triplet correlations for a canted-AF/S bilayer have been plotted as a function of space for filling factor $f=0.6$. We see that $F^{t,z}$, the component of $\vec{F^t}$ perpendicular to the canted-AF sublattice magnetizations plane, is also being generated along with the in-plane components $F^{t,x}$ and $F^{t,y}$. It oscillates from a constant positive to negative value with N\'{e}el order. \ak{This oscillation can be understood in terms of the noncollinearity that generates $F^{t,z}$. From one lattice site to the next, the angle between the spin-splitting at the site and its direct neighbors is changing sign.} However, $F^{t,z}$ only appears for non-zero $\mu$, and is zero at half-filling ($f=0.5$) where $\mu=0$. $F^{t,x}$ too oscillates between a constant positive and negative value with N\'{e}el order, while $F^{t,y}$ is constant in space. Both $F^{t,x}$ and $F^{t,y}$ are imaginary, consistent with previous theoretical results for F/S bilayers~\cite{Champel2005, Eschrig2007}. The AF sublattice magnetization configuration has been so chosen here to obtain and focus on the non-zero and spatially constant sum of the equal-spin triplets $F_{\uparrow\uparrow}$ and $F_{\downarrow\downarrow}$. In Appendix~\ref{axischange}, we present the correlations for a configuration when the AF N\'eel order is aligned with the $z$-axis~\cite{Bobkov2022}. To conclude this discussion, the intrinsic noncollinearity of the canted AF successfully generates all three components of the spin-triplet correlations. \ak{The generation of spinful triplet Cooper pairs has conventionally been accomplished via the noncollinearity between the magnetizations of different ferromagnetic layers~\cite{Leksin2012}.}

\begin{figure*}[tb]
	\begin{center} 
		\includegraphics[width=\linewidth]{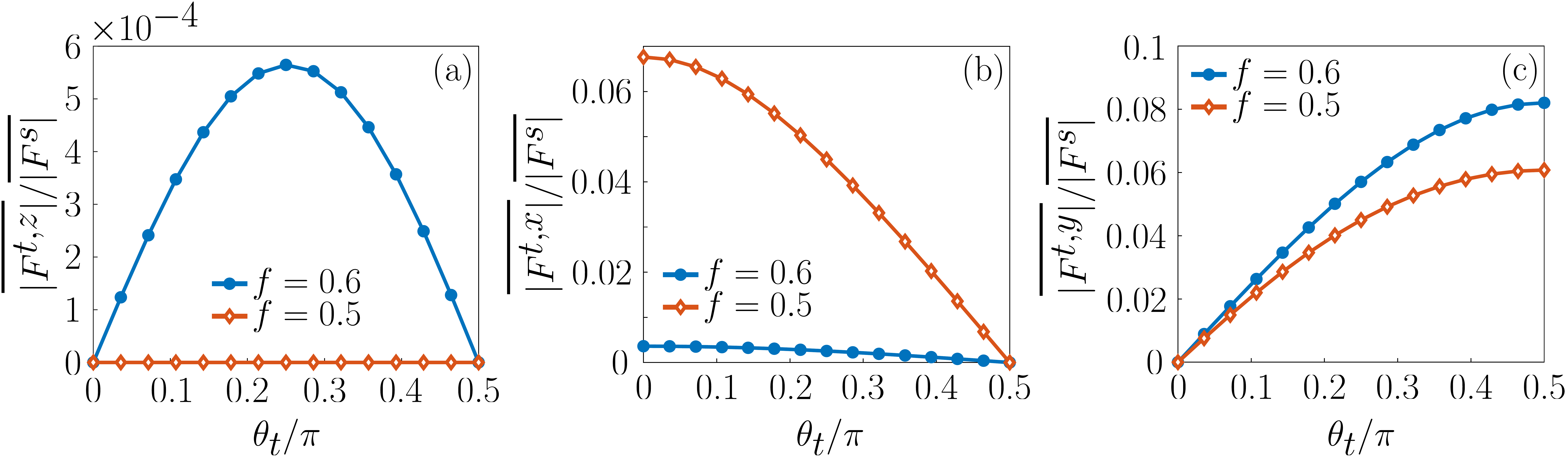}
		\caption{Variation of triplet correlations with canting angle $\theta_t$ for filling factors $f=0.5$ $(\mu=0)$ and $f=0.6$ $(\mu/\Delta_0 \approx 37)$, where $\Delta_0$ represents the zero-temperature value of the superconducting order parameter in absence of the canted-AF layer. (a) The average magnitude of the normalized spin-triplet correlation $F^{t,z}$ is maximum when the noncollinearity between the two sublattices is maximum, i.e. at $\theta_t=\pi/4$. However, it is zero at half-filling ($f=0.5$). (b) Average magnitude of the normalized spin-triplet correlation $F^{t,x}=[F_{\uparrow\uparrow}-F_{\downarrow\downarrow}]/2$ decreases as the effective antiferromagnetism becomes weaker with increasing $\theta_t$. (c) The average magnitude of the normalized spin-triplet correlation $F^{t,y}=[i(F_{\uparrow\uparrow}+F_{\downarrow\downarrow})]/2$ increases as the effective ferromagnetism becomes stronger with increasing $\theta_t$. The averages are taken over all the sites and are denoted via an overhead bar. The detailed parameters employed for the numerical evaluation are specified in Appendix~\ref{parameter}.}
		\label{fig:3}
	\end{center}
\end{figure*}

We now examine the dependence of these spin-triplets on the canting angle. To this end, we plot the average magnitudes of the three spin-triplet correlations vs.~the canting angle $\theta_t$ in Fig.~\ref{fig:3}. As we change $\theta_t$ from $0$ to $\pi/2$ the system changes from a collinear AF (along $x$-axis) to a collinear F (along $y$-axis). As discussed above, $F^{t,z}$ is found to vanish identically at $\mu=0$ ($f=0.5$) for all canting angles. This will be explained further below. However, for non-zero $\mu$ (away from the half-filling case), it increases from 0 to a finite value as we go from a collinear antiferromagnetic alignment to maximal noncollinearity between the sublattice magnetic moments, and decreases back to zero in the ferromagnetic alignment [Fig.~\ref{fig:3}(a)]. This component thus results directly due to the AF sublattice magnetization noncollinearity. 

The spatially averaged value of $|F^{t,x}|$ decreases as $\theta_t$ goes from 0 to $\pi/2$ [Fig.~\ref{fig:3}(b)]. This component, therefore, essentially follows the N\'eel vector magnitude and appears to stem directly from the antiferromagnetism~\cite{Bobkov2022}. It has been understood as being due to the interband pairing, which is feasible when $\mu$ is smaller or comparable to the superconducting gap. However, we also find such Néel triplets to be present for $\mu \approx 37\Delta_0$ ($f=0.6$), although they are significantly weaker than for the case of $\mu=0$ [Fig.~\ref{fig:3}(b)]. We attribute this observation to a modification of the normal state electronic wavefunctions by the AF, so that even the conventional intraband pairing causes a finite generation of the N\'eel triplets. Finally, the average value of $|F^{t,y}|$ increases with the canting angle and it appears to be caused primarily by the net magnetization.

\subsection{Insights from simplified analytics}
\label{sec:decoupling}

In order to understand the difference between $\mu=0$ and $\mu\neq 0$ cases, we examine the electronic properties of a bilayer comprising a normal metal and an AF, as detailed in Appendix~\ref{AFNsection}. For $\mu=0$, the Fermi wave-vector is $k_F=\pi/2a$. This means that the electrons participating in the formation of Cooper pairs have $|k|\sim \pi/2a$. The eigenfunctions with $|k|\sim \pi/2a$ are such that the probability of finding an electron is non-zero on one sublattice while it is zero on the other sublattice. Now, the electrons near the Fermi level of the superconductor which interact with sublattice A do not see sublattice B and vice-versa. Therefore the triplet correlations are generated independently by sublattices A and B. The resultant correlations we obtain are a sum of correlations generated by the two sublattices. So, the only non-zero components are $F^{t,x}$ and $F^{t,y}$. This separation of the two sublattices at a special value of the electronic chemical potential is reminiscent of a similar result obtained for spin pumping via AFs~\cite{Takei2014,Cheng2014,Kamra2017,Liu2017}.

In contrast, for $\mu \neq 0$, the Fermi level is within one of the bands and the wavefunctions of the states near the Fermi energy are such that the electrons on a site of the A sublattice also have a non-zero probability at the sites of the B sublattice. There is no way for the electrons to arrange themselves to decouple the two sublattices~\cite{Kamra2017,Liu2017}. The electrons experience a spin-splitting field in one direction at site 1 (of sublattice A) and in another direction at site 2 (of sublattice B), then again the first orientation at site 3 (of sublattice A). So, the electrons see the noncollinearity between the magnetic moments on the adjacent lattice sites. Therefore, the correlation $F^{t,z}$ along the direction perpendicular to the plane of magnetic moments of the canted-AF becomes non-zero along with the in-plane components $F^{t,x}$ and $F^{t,y}$ when $\mu \neq 0$. At the same time, the electronic amplitudes at the two sublattices are different due to the adjacent AF, as detailed in Appendix~\ref{AFNsection}. This lends the normal electronic states a weak N\'eel character which manifests itself in the emergence of N\'eel triplets even for the conventional intraband pairing.


\section{Critical temperature}\label{sec:Tc}

The formation of spin-triplet Cooper pairs comes at the cost of destroying their spin-singlet counterparts that are originally produced in and stabilize the superconducting state~\cite{Chandrasekhar1962,Clogston1962,Maki1964}. Hence, the critical temperature is reduced with the formation of spin-triplets, which may offer a convenient experimental signature. Thus, we investigate the critical temperature of our AF/S bilayer now via numerical self-consistent solution of the BdG equation \eqref{eq1}.

\begin{figure}[tb]
	\begin{center} 
		\includegraphics[width=140mm]{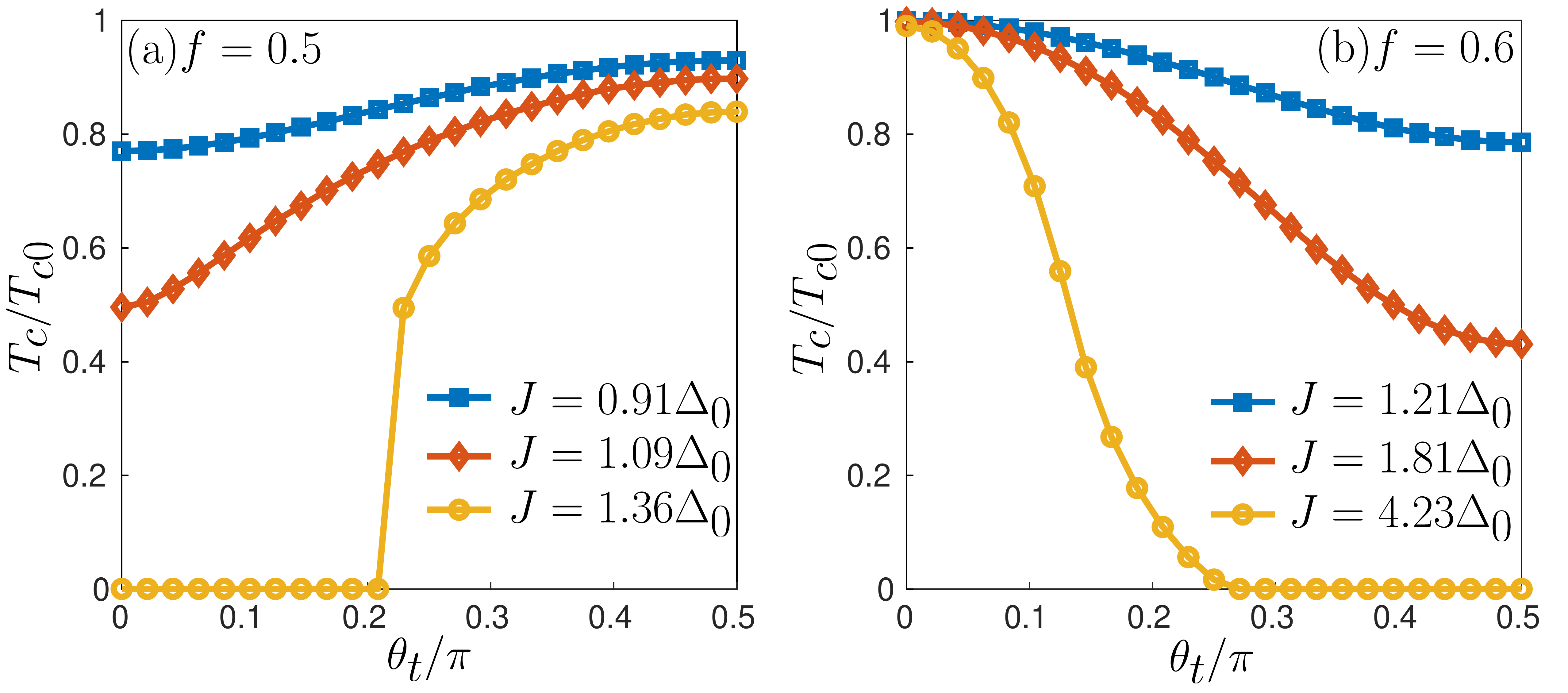}
		\caption{(a) Normalized critical temperature $T_c$ vs.~canting angle $\theta_t$ for filling factor (a) $f=0.5$ and (b) $f = 0.6$ considering different values of the spin-splitting $J$. Here, $T_{c0}$ and $2\Delta_0$ are respectively the critical temperature and the zero-temperature superconducting gap of the same superconductor without the AF layer. The detailed parameters employed for the numerical evaluation are specified in Appendix~\ref{parameter}.}
		\label{fig:5}
	\end{center}
\end{figure}

Critical temperature $T_c$ vs.~canting angle $\theta_t$ is plotted in Fig.~\ref{fig:5} for (a) $f = 0.5$ ($\mu=0$) and (b) $f = 0.6$ ($\mu \approx 37\Delta_0$). We find that for $\mu=0$, $T_c$ increases with $\theta_t$ while it manifests the opposite dependence for $\mu \neq 0$. These intriguing and distinct dependencies can be understood based on our analysis of spin-triplets generation above. 

Let us first consider the $f = 0.5$, corresponding to $\mu = 0$, case presented in Fig.~\ref{fig:5}(a). In this case, a strong generation of spin-zero N\'eel triplets [Fig.~\ref{fig:3}(b)] due to interband pairing leads to maximal $T_c$ suppression at $\theta_t = 0$. Hence, the $T_c$ increases with $\theta_t$ since the $T_c$ suppression is stronger for the collinear AF case ($\theta_t = 0$) than for the F case ($\theta_t = \pi/2$). Further, when the exchange field is large enough, we find a complete suppression of superconductivity at $\theta_t = 0$ corresponding to a vanishing $T_c$. The abrupt change in $T_c$ with $\theta_t$ here is attributed to the additional contribution to superconductivity suppression by the opening of a normal dispersion bandgap by the AF, as described in Appendix~\ref{AFNsection}. This normal state bandgap predominantly affects the superconducting pairing at half-filling when $\mu = 0$.

For the case of $f = 0.6$, the N\'eel spin-triplets generation by the antiferromagnetic order is much weaker (Fig.~\ref{fig:3}). On the other hand, the ordinary spin-triplets generation by a ferromagnet remain of the same order of magnitude as for $f = 0.5$. Thus, $T_c$ is largest for $\theta_t = 0$ and it decreases with $\theta_t$. The amplitude of spin-triplets generated due to the noncollinearity [Fig.~\ref{fig:3}(a)] remains small and does not seem to affect the $T_c$ dependence substantially. Contrary to the $\mu=0$ case, the variation of $T_c$ with $\theta_t$ is smooth even for large values of the exchange field $J$.   


\section{Correlations in 2-D}\label{sec:2D}

In our discussion above, we have considered a one-dimensional (1-D) superconductor with the aim of examining essential physics employing analytic results discussed in Appendix~\ref{AFNsection}. We now validate these results using a two-dimensional (2-D) model for the S. This further allows us to examine how the spin-triplet correlations vary with space as we move away from the S/AF interface. 

In Fig.~\ref{fig:1}(b), the superconducting lattice is along the $x$-axis. We add more such 1-D layers in the $y$-direction to create our 2-D model for the S. Each site of this 2-D sheet is indexed as $(j_x,j_y)$ so that $j_x$ takes a value between 1 to $N_x$, and $j_y$ takes a value between 1 to $N_y$, where $N_x$ and $N_y$ are the number of sites along $x$- and $y$-direction, respectively. The spin-splitting effect is experienced only by the electrons at the sites next to the AF/S interface. Thus, it suffices to treat the AF via the same 1-D model as before. Overall, the Hamiltonian  of Eq.~\eqref{eq1} is modified by letting all site indices $j$ take the form $(j_x,j_y)$. Summation in the spin-splitting term is now over the sites with indices of the form $(j_x,1)$. We continue to consider periodic boundary condition along the $x$-axis, like in the 1-D case.

\begin{figure*}[tb]
	\begin{center} 
		\includegraphics[width=\linewidth]{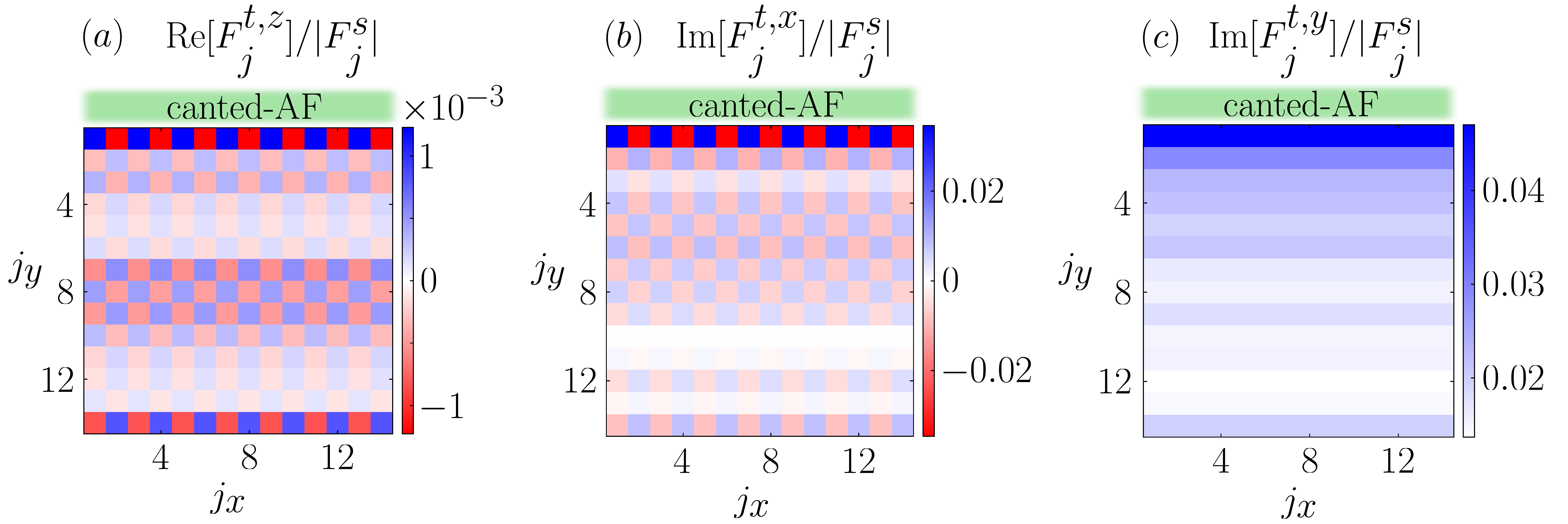}
		\caption{Spatial variation of the spin-triplet correlations for a 2-D superconductor. The sites are indexed as $(j_x, j_y)$. The sites with index $(j_x, 1)$ form the layer adjacent to the AF. A small section of size $14\times14$ of a superconducting sheet of size $102 \times 14$ has been plotted here for clarity. The real part of the normalized zero-spin triplet correlation $F^{t,z}_j$ [panel (a)], and the imaginary parts of the normalized spin-triplet correlations $F^{t,x}_j$ [panel (b)] and $F^{t,y}_j$ [panel (c)] have been plotted as colormaps. The imaginary part of the former and the real parts of the latter two are zero. We consider maximal noncollinearity ($\theta_t = \pi/4$) and filling fraction $f=0.6$ \add{($\mu = 18.5 \overline{|\Delta_0|}$). Here, $\overline{|\Delta_0|}$ is the average magnitude of the superconducting order parameter in absence of the canted-AF layer} and all the correlations are normalized with respect to the magnitude of the singlet correlation $|F^s_j|$. The detailed parameters employed for the numerical evaluation are specified in Appendix~\ref{parameter}.}
		\label{fig:4}
	\end{center}
\end{figure*}

Carrying out the BdG diagonalization self-consistently and numerically, we evaluate the spatially-resolved spin-triplet correlations [Eqs.~\eqref{corr2}-\eqref{corr5}] for this system and plot them in Fig.~\ref{fig:4}. We have the same observations in the first layer (along the AF/S interface) of the 2-D sheet as the 1-D case discussed in section \ref{Tripcorr}. Additional calculations not presented here confirm that $F^{t,z}$ appears only when $\mu$ is non-zero. We observe that $F^{t,x}$ and $F^{t,z}$ show modulation between positive and negative values along the $y$-axis apart from along the $x$-axis. This is interesting because the system has nothing imposing N\'eel order along the $y$-axis on the correlations. 
In addition, we find some layers in which the alternating pattern of positive and negative values is skipped. This is attributed to Friedel-like oscillations~\cite{Zhu2016,Bobkov2022}. We found that these skipping of pattern only appears for non-zero $\mu$ while we get a perfect alternating pattern along the $y$-axis for $\mu=0$ case. \add{Although these N\'{e}el triplet correlations flip sign at the length scale of a lattice constant, their magnitude decays at the coherence length scale (shown in Appendix~\ref{corrndecay}).} The correlation component $F^{t,y}$ [Fig.~\ref{fig:4}(c)] is constant in each layer along the interface (along the $x$-axis) but decays as we move away from the interface along the $y$-direction. So, the 2-D case is consistent with and corroborates our 1-D results, \add{resulting in a similar trend in $T_c$ suppression (Appendix~\ref{Tc2D}).} We find that the spatial pattern (oscillating with N\'eel order or being constant in space) imposed on superconducting correlations along the interfacial direction also manifests itself perpendicular to the AF/S interface.

\begin{figure*}[tb]
	\begin{center} 
		\includegraphics[width=\linewidth]{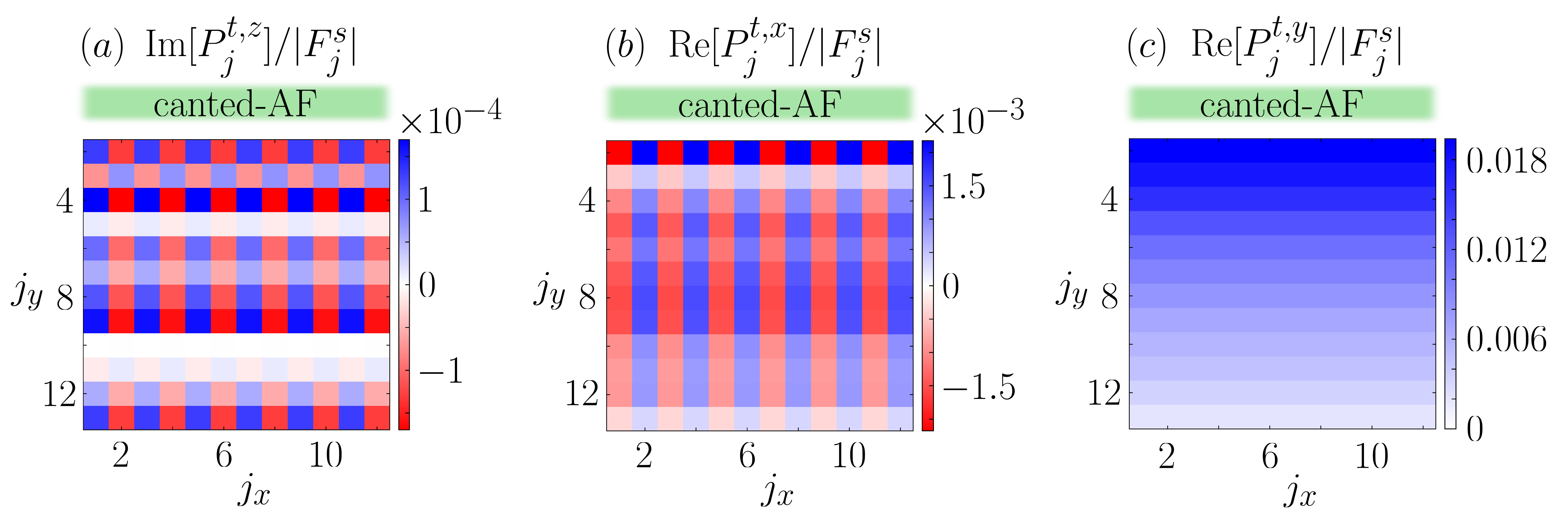}
		\caption{\lina{Spatial variation of the $p_y$-wave triplet correlations in a 2-D superconductor. The sites are indexed as $(j_x, j_y)$. The sites with index $(j_x, 1)$ form the layer adjacent to the AF. A small section of size $12\times12$ of a superconducting sheet of size $202 \times 14$ has been plotted here for clarity. The imaginary part of the normalized zero-spin $p_y$-wave triplet correlation $P^{t,z}_j$ [panel~(a)], and the real parts of the normalized $p_y$-wave spin-triplet correlations $P^{t,x}_j$ [panel~(b)] and $P^{t,y}_j$ [panel~(c)] have been plotted here. The real part of the former and the imaginary parts of the latter two are zero. We consider maximal noncollinearity ($\theta_t = \pi/4$) and filling factor $f=0.6$. All the correlations are normalized with respect to the magnitude of the $s$-wave singlet correlation $|F^s_j|$. The detailed parameters employed for the numerical evaluation are specified in Appendix~\ref{parameter}.}}
		\label{fig:pycorrn}
	\end{center}
\end{figure*}

\lina{In the 2-D case, where the AF/S interface breaks translational symmetry along the $y$-axis, even-frequency $p_y$-wave spin-triplet correlations are created, in addition to the odd-frequency $s$-wave ones \cite{Tanaka2012,Linder2011}. These are \ak{quantified as~\cite{Olthof2021}}}
\add{
\begin{align}
P^{t,z}_j = \frac{1}{8} \left[ \langle c_{j,\downarrow} c_{j+\hat{y},\uparrow} \rangle + \langle c_{j,\uparrow} c_{j+\hat{y},\downarrow} \rangle - \langle c_{j,\downarrow} c_{j-\hat{y},\uparrow} \rangle - \langle c_{j,\uparrow} c_{j-\hat{y},\downarrow} \rangle  \right],\\
P^{t,x}_j = \frac{1}{8} \left[ -\langle c_{j,\uparrow} c_{j+\hat{y},\uparrow} \rangle + \langle c_{j,\downarrow} c_{j+\hat{y},\downarrow} \rangle + \langle c_{j,\uparrow} c_{j-\hat{y},\uparrow} \rangle - \langle c_{j,\downarrow} c_{j-\hat{y},\downarrow} \rangle \right],\label{eq:Px}\\
P^{t,y}_j = \frac{i}{8} \left[ -\langle c_{j,\uparrow} c_{j+\hat{y},\uparrow} \rangle - \langle c_{j,\downarrow} c_{j+\hat{y},\downarrow} \rangle + \langle c_{j,\uparrow} c_{j-\hat{y},\uparrow} \rangle + \langle c_{j,\downarrow} c_{j-\hat{y},\downarrow} \rangle \right],
\end{align}
where the site index $j\pm \hat{y} \equiv (j_x,j_y \pm 1)$ are the nearest neighbors of site $j$ along the $y$-axis, and $P^{t,\alpha}_j$ represent the zero-spin $p_y$-wave triplet correlation when the spin is measured along the $\alpha$-axis ($\alpha=x,y,z$).}

\lina{While the $s$-wave spin-triplets in Fig.~\ref{fig:4}(a)-(b) nearly follow a checkerboard pattern, $P^{t,z}$ and $P^{t,x}$ tend to form stripes along the $y$-direction [Fig.~\ref{fig:pycorrn}(a)-(b)]~\footnote{As the $p_y$-wave triplet correlations are not defined at the edges, we plot these starting at the second row of lattice sites from the edge.}. Unlike the on-site $s$-wave pairing, the $p_y$-wave pairing cannot take advantage of the complete decoupling of the two sublattices at half-filling as they are defined on the link between sites in the A and B sublattice. For $P^{t,x}$, this can be seen by realizing that $\langle c_{j,\sigma}c_{j+\hat{y},\sigma}\rangle=-\langle c_{(j+\hat{y}),\sigma}c_{(j+\hat{y})-\hat{y},\sigma}\rangle$. Thus, if the link in the positive $y$-direction from site $j$ gives a positive contribution to $P^{t,x}$, then the link in the negative $y$-direction from site $j+\hat{y}$ also gives a positive contribution, as can be seen by insertion into Eq.~\eqref{eq:Px}. Thus, a link from the A sublattice to the B sublattice and a link from the B sublattice to the A sublattice give a contribution to $P^{t,x}$ of the same sign. At half-filling, the sign of $P^{t,x}$ is thus dictated by the sign of the spin-splitting induced at the lattice site closest to the interface. Away from half-filling, where the two sublattices are no longer perfectly decoupled, the perfect stripe pattern can be shifted along the $x$-direction as can be seen for $P^{t,z}$ in Fig.~\ref{fig:pycorrn}(a). The ferromagnetic component $P_j^{t,y}$ in Fig.~\ref{fig:pycorrn}(c) remains positive inside the whole superconductor, similar to the $s$-wave triplets in Fig.~\ref{fig:4}(c). Although we here observe qualitatively different behavior in the nearest-neighbor \ak{(p-wave)} pairing compared to the on-site \ak{(s-wave)} one, we must note that the antiferromagnetic component and the component resulting from canting are both one order of magnitude larger for the latter. \ak{Thus, the AF's influence on the critical temperature and other key superconducting properties is dominated by the odd-frequency s-wave N\'eel triplets generated in the S.}}


\section{Concluding remarks}\label{sec:summary}

We have theoretically demonstrated the generation of all, including the spinful, spin-triplet Cooper pairs in a conventional superconductor by an adjacent canted antiferromagnet. Our proposal leverages the intrinsic noncollinearity between the two sublattice magnetizations in a canted antiferromagnet for applications in superconducting hybrids. This canting can be induced intrinsically by Dzyaloshinskii-Moriya interaction, such as in hematite~\cite{Morrish1995}. Additionally, it can be induced and controlled using an applied magnetic field~\cite{Wimmer2020}. The resulting spin-triplets have a predominantly N\'eel character, i.e., their amplitude oscillates in space on the lattice length scale similar to the N\'eel spin order. \ak{While we have considered a lattice matched interface for concreteness and simplicity, the essential physics remains the same even in the presence of disorder with a gradual suppression of the N\'eel triplets with increasing disorder~\cite{Bobkov2022}. Furthermore, a metallic AF would lead to similar phenomena as discussed here, with the additional complication of Cooper pairs leaking into the AF.} The superconducting critical temperature is more strongly suppressed by the interband Néel spin-triplets than by a ferromagnetic spin-splitting field of similar magnitude, thereby offering an experimental signature of their generation. Altogether, our analysis highlights the noncollinear nature of homogeneous canted antiferromagnets by employing them for generating spin-triplet Cooper pairs in a simple superconducting bilayer. This manner of generating noncollinearity using a homogeneous antiferromagnet is expected to find use in other phenomena that have traditionally relied on magnetic multilayers or spin textures.


\appendix

\section{Normal metal interfaced with an antiferromagnet} \label{AFNsection}
In order to understand when interband and intraband pairing is favored, we  consider the normal-state wave functions of an AF/normal metal bilayer. The Hamiltonian for the conducting electrons of a normal metal interfaced with an insulating AF is modeled as~\cite{Simensen2020}
\begin{align} \label{AFNham}
	H = &  -\mu \sum_{j,\sigma} c_{j,\sigma}^\dagger c_{j,\sigma} -t \sum_{\left< i,j\right>}\sum_{\sigma} c_{i,\sigma}^\dagger c_{j,\sigma}-\frac{J}{2} \sum_{j} \vec{M}_j \cdot \vec{S}_j,
\end{align}
where $\vec{M_j}=(-1)^{j+1} \hat{z}$ gives the magnetic texture of a collinear AF. Other symbols have the same meaning as in section \ref{model} of the main text.

In order to calculate the eigenenergies and eigenvectors, the Hamiltonian is written in terms of creation and annihilation operators for electrons at sublattices A and B. Creation operators for electrons at sublattices A and B are defined as $a_{j,\sigma}^\dagger=c^\dagger_{2j-1,\sigma}$, and $b_{j,\sigma}^\dagger=c^\dagger_{2j,\sigma}$, where $j=1,2,\dots,N/2$. Then, the basis is changed from Wannier wavefunctions to Bloch wavefunctions using the relation
\begin{align}
	a^\dagger_{j,\sigma}=\sum_{k\in FBZ} \frac{1}{\sqrt{N/2}} e^{-ik(2j-2)a} a_{k,\sigma}^\dagger, \quad \text{and} \quad
	b_{j,\sigma}^\dagger=\sum_{k\in FBZ} \frac{1}{\sqrt{N/2}} e^{-ik(2j-1)a} b_{k,\sigma}^\dagger,
\end{align}
where $a$ is the lattice constant, and $k$ is a reciprocal lattice vector in the first Brillouin zone (FBZ). Now, the Hamiltonian can be written in the form
\begin{align} \label{AFNham2}
	H=\sum_k \begin{pmatrix}
		a_{k,\uparrow}^\dagger & a_{k,\downarrow}^\dagger & b_{k,\uparrow}^\dagger & b_{k,\downarrow}^\dagger
	\end{pmatrix} H(k) \begin{pmatrix}
		a_{k,\uparrow} & a_{k,\downarrow} & b_{k,\uparrow} & b_{k,\downarrow}
	\end{pmatrix}^T,
\end{align}
giving the dispersion relation
\begin{align}
	E_\pm(k)=-\mu \pm \sqrt{4t^2 \cos^2(ka)+(J/2)^2},
\end{align}
where the two energy bands $E_\pm(k)$ are two-fold degenerate. The eigenvectors for energy band $E_-(k)$ are 
\begin{align} \label{psi3}
\psi_1(k)=&\mathcal{N}_2\begin{pmatrix}
2t\cos(ka)\\
0\\
-J/2+\sqrt{4t^2\cos^2(ka)+(J/2)^2}\\
0
\end{pmatrix} =  \begin{pmatrix}
u_{Ak\uparrow}^{(1)}\\
u_{Ak\downarrow}^{(1)}\\
u_{Bk\uparrow}^{(1)}\\
u_{Bk\downarrow}^{(1)}
\end{pmatrix},  \text{ and} \\
\psi_2(k)=&\mathcal{N}_1\begin{pmatrix}
0\\
2t\cos(ka)\\
0\\
J/2+\sqrt{4t^2\cos^2(ka)+(J/2)^2}
\end{pmatrix} =  \begin{pmatrix}
u_{Ak\uparrow}^{(2)}\\
u_{Ak\downarrow}^{(2)}\\
u_{Bk\uparrow}^{(2)}\\
u_{Bk\downarrow}^{(2)}
\end{pmatrix}.  \label{psi4}
\end{align}

Whereas the eigenvectors for the energy band $E_{+}(k)$ are
\begin{align} \label{psi1}
\psi_3(k)=& \mathcal{N}_1\begin{pmatrix}
2t\cos(ka)\\
0\\
-J/2-\sqrt{4t^2\cos^2(ka)+(J/2)^2}\\
0
\end{pmatrix} =  \begin{pmatrix}
u_{Ak\uparrow}^{(3)}\\
u_{Ak\downarrow}^{(3)}\\
u_{Bk\uparrow}^{(3)}\\
u_{Bk\downarrow}^{(3)}
\end{pmatrix}, \text{ and} \\
\psi_4(k)=&\mathcal{N}_2\begin{pmatrix}
0\\
2t\cos(ka)\\
0\\
J/2-\sqrt{4t^2\cos^2(ka)+(J/2)^2}
\end{pmatrix} =  \begin{pmatrix}
u_{Ak\uparrow}^{(4)}\\
u_{Ak\downarrow}^{(4)}\\
u_{Bk\uparrow}^{(4)}\\
u_{Bk\downarrow}^{(4)}
\end{pmatrix}.  \label{psi2}
\end{align}

\begin{figure*}[tb]
	\begin{center} 
		\includegraphics[width=0.5\linewidth]{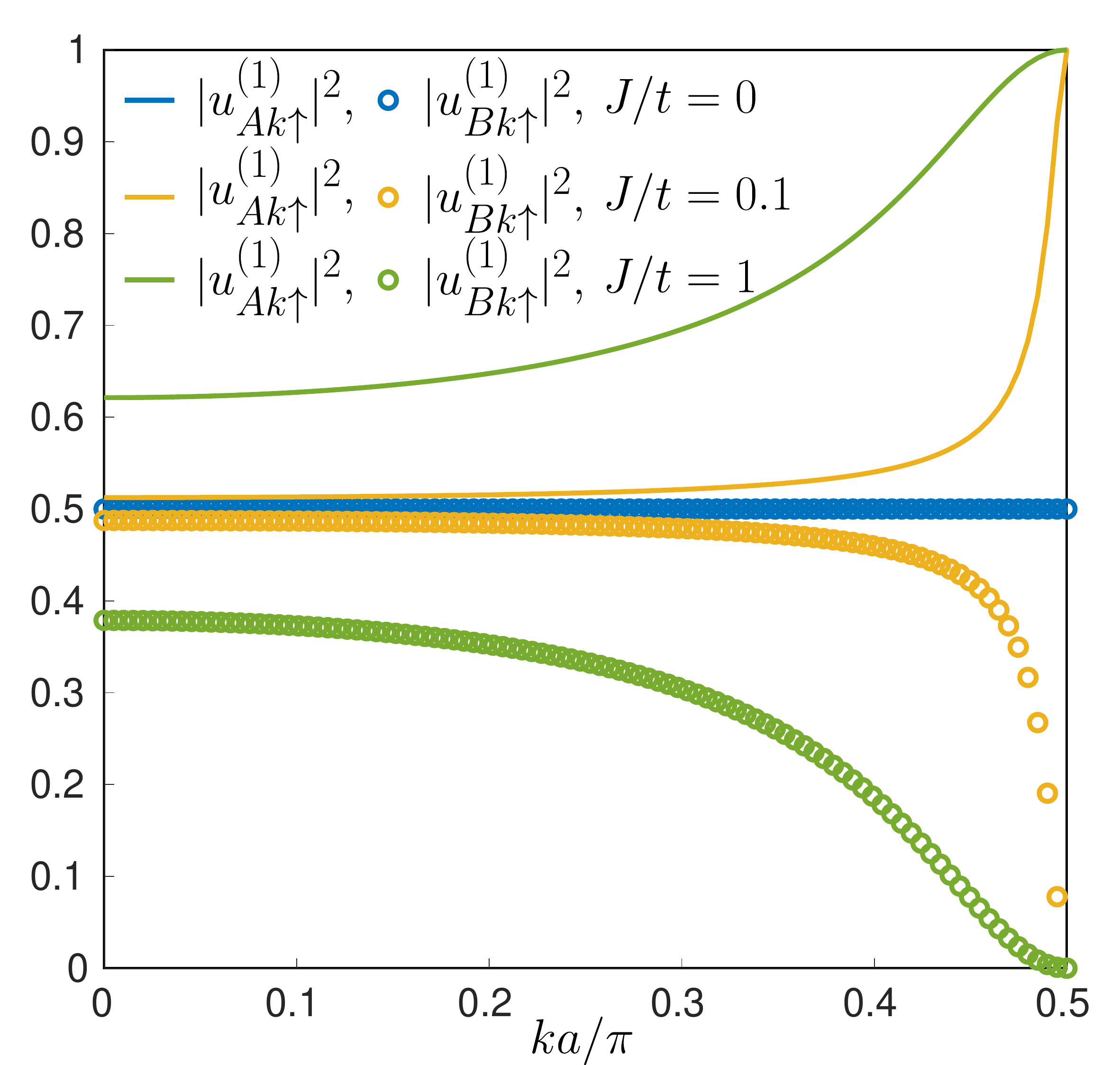}
		\caption{Non-zero components of eigenfunction $\psi_1$ (Eq.~\eqref{psi3}) of a normal metal/AF bilayer for different strengths of magnetic exchange interaction as a function of positive $k$ values in the first Brillouin zone.}
		\label{AFNpsi}
	\end{center}
\end{figure*}

Here, $\mathcal{N}_1$ and $\mathcal{N}_2$ are the normalization factors of the eigenvectors. $\psi_1$ and $\psi_3$ correspond to the wavefunctions of spin $\uparrow$ electrons, whereas $\psi_2$ and $\psi_4$ correspond to spin $\downarrow$ electrons. Fig.~\ref{AFNpsi} shows a plot of the non-zero components of $\psi_1$ for positive $k$ values . Let us compare the case of $J=0$ and $J\neq 0$ to study how the states of a metal are modified when it is brought in contact with an AF.

From Fig.~\ref{AFNpsi}, we see that the probabilities of finding a spin $\uparrow$ electron at sites of sublattice A and B are equal for $J=0$. However, for $J\neq0$, the probability on A sublattice is more than that on B sublattice. It is important to note that the probability of finding spin $\uparrow$ electrons at sublattice A sites becomes 1 and that for sublattice B becomes 0 at $k=\pi/2a$. Similar asymmetries in the sublattices arise for states $\psi_2$, $\psi_3$, and $\psi_4$ as soon as we make $J$ non-zero. For $\psi_2$, the probability of finding a spin $\downarrow$ electron at sublattice site B becomes more than that for sublattice A. For the states of $E_+(k)$ band ($\psi_3$ and $\psi_4$), the roles of sublattices A and B are interchanged with respect to spin.

Near the Brillouin zone boundary (BZB) of band $E_-(k)$ (i.e. $\psi_1$ and $\psi_2$), all the spin $\uparrow$ electrons get localized to the sites of sublattice A and all the spin $\downarrow$ electrons get localized to the sites of sublattice B. For the band $E_+(k)$, sublattices A and B interchange their roles and we find that spin $\uparrow$ electrons get localized at sublattice B and spin $\downarrow$ electrons get localized at sublattice A near the BZB. This property of the electronic states near the BZB is the reason why interband pairing is the dominant mechanism for the formation of on-site opposite-spin Cooper pairs in an AF/S bilayer when the Fermi level lies within the antiferromagnetic band gap $E_+ - E_-$. On the other hand, intraband pairing is the dominant mechanism when the Fermi level lies within one of the two bands.


\section{Bogoliubov-de Gennes calculation} \label{AC}

The Hamiltonian in Eq.~\eqref{eq1} can be written as
\begin{align}
H =-\mu N + \sum_j \frac{|\Delta_j|^2}{U}+\frac{1}{2}\sum_{i,j} \Psi_i^\dagger \tilde{H}_{i,j} \Psi_j,
\end{align}
where $\Psi_j^\dagger = \begin{pmatrix}
c_{j,\uparrow}^\dagger & c_{j,\downarrow}^\dagger & c_{j,\uparrow} & c_{j,\downarrow}
\end{pmatrix}$ and $N$ is the total number of sites. The matrix $\Tilde{H}$ is diagonalized by solving the BdG equations
\begin{align} 
 \sum_j \tilde{H}_{i,j}\phi_{j,n}=E_n \phi_{i,n}, \text{ where} \\
  \phi_{i,n} = \begin{pmatrix}
u_{i,n,\uparrow} & u_{i,n,\downarrow} & v_{i,n,\uparrow} & v_{i,n,\downarrow} \end{pmatrix}^T,
\end{align}
is the eigenvector and $E_n$ the eigenenergies of $\tilde{H}$. Now the Hamiltonian can be written as
\begin{align}
H=-\mu N + \sum_j \frac{|\Delta_j|^2}{U} - \frac{1}{2} \tilde{\sum_n} E_n + \tilde{\sum_n} \Gamma_n^\dagger E_n \Gamma_n, 
\end{align}
where $\Tilde{\sum}_n$ represents sum over positive eigenenergies, and $\Gamma_n$'s are Bogoliubov fermionic operators related to the old fermionic operators by $c_{j,\sigma}=\tilde{\sum}_n(u_{j,n,\sigma}\Gamma_n+v_{j,n,\sigma}^* \Gamma_n^\dagger)$. The superconducting order parameter is calculated self-consistently using the relation
\begin{align}
\Delta_j=-U\tilde{\sum_n} \left[u_{j,n,\downarrow} v_{j,n,\uparrow}^* [1-f_{FD}(E_n)]+v_{j,n,\downarrow}^* u_{j,n,\uparrow} f_{FD}(E_n)\right],
\end{align}
where $f_{FD}(E_n)=\langle \Gamma_n^\dagger \Gamma_n\rangle=1/(e^{E_n/k_B T}+1)$ is the Fermi-Dirac distribution. 

In order to calculate the spin-triplet correlations, we consider the anomalous Matsubara Green's function $F_{jj,\sigma\sigma'}(\tau)=-\langle T_\tau c_{j,\sigma}(\tau) c_{j,\sigma'}(0) \rangle$, where $\tau=i\tilde{t}$ is the imaginary time, $\tilde{t}$ is the time, and $T_\tau$ is the ordering operator for $\tau$. Taking its Fourier transform, we get
\begin{align}
F_{jj,\sigma\sigma'}(i\omega_l)=\int_0^{\beta} e^{i\omega_l \tau} F_{jj,\sigma\sigma'}(\tau) \, d\tau = \tilde{\sum_n} \left[ \frac{u_{j,n,\sigma}v_{j,n,\sigma'}^*}{i\omega_l-E_n/\hbar}+\frac{v_{j,n,\sigma}^*u_{j,n,\sigma'}}{i\omega_l+E_n/\hbar}\right],
\end{align}
where $\beta=\hbar/k_B T$, $k_B$ is the Boltzmann constant, $T$ is the temperature, and $\omega_l=(2l+1)\pi/\beta$ is a Matsubara frequency for fermions with integer $l$. This expression is used in calculating the correlations in Eq.~\eqref{corr1} and their relevant components Eqs.~\eqref{corr2}-~\eqref{corr5}. 

\add{

\section{Decay of triplet correlations with distance in 2-D} \label{corrndecay}

\begin{figure*}[tb]
	\begin{center} 
		\includegraphics[width=0.5\linewidth]{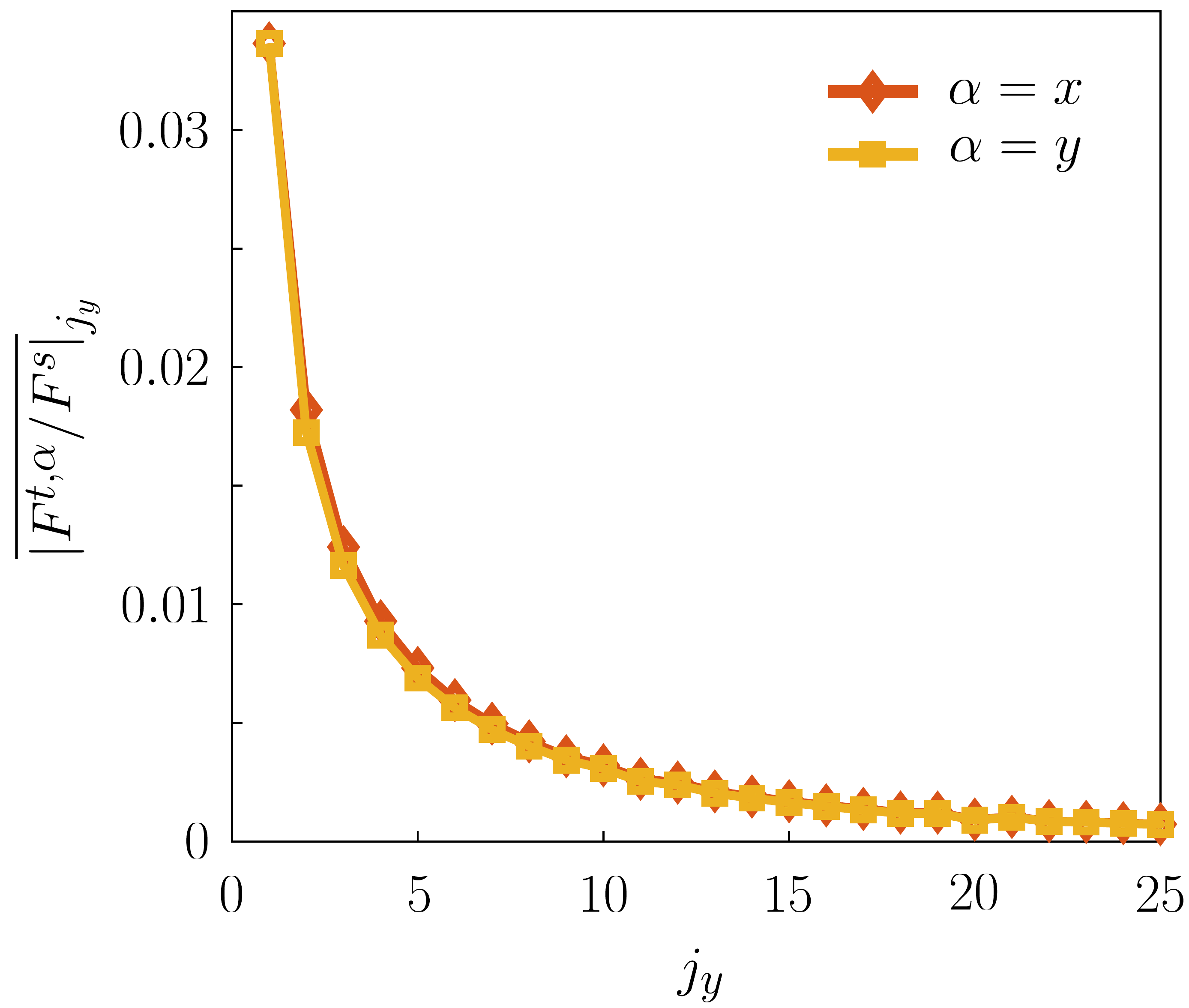}
		\caption{\lina{The average magnitude of the $s$-wave spin-triplet correlations normalized by the singlet correlation in each $j_y$ layer parallel to the canted-AF/S interface $\left(\overline{|F^{t,\alpha}/F^s|}_{j_y}, \,\, \alpha=x,y\right)$ is plotted with respect to the distance from the interface. We consider half filling ($f=0.5$) and maximal canting ($\theta_t=\pi/4$). The sites with index $j_y=1$ form the layer adjacent to the AF.} Ignoring the edge effects, we show only the first $25$ layers of a superconducting sheet with $102\times30$ sites. \lina{The $F^{t,z}$ component is zero in all the layers.} The detailed parameters employed for the numerical evaluation are specified in Appendix~\ref{parameter}.}
		\label{fig:corr_vs_jy}
	\end{center}
\end{figure*} 

In section~\ref{sec:2D}, we discussed the spatial variation of triplet correlations in a 2-D superconductor interfaced with a canted-AF and observed that the N\'eel triplet correlations flip sign on adjacent sites to form a checkerboard pattern. 
\lina{In Fig.~\ref{fig:corr_vs_jy}, we study the decay of the $s$-wave spin-triplet correlations with the distance from the AF/S interface. We consider half-filling ($f=0.5$) to avoid the influence of Friedel oscillations. We observe that the magnitude of both the ferromagnetic and antiferromagnetic components of the correlations decay exponentially over a length scale determined by the coherence length. The rapid oscillations of the N\'eel order of the canted-AF over the atomic length scale thus only affects the sign of the triplet correlations.}


\section{\add{Critical temperature of 2-D S/canted-AF}} \label{Tc2D}

\begin{figure*}[tb]
	\begin{center} 
		\includegraphics[width=0.5\linewidth]{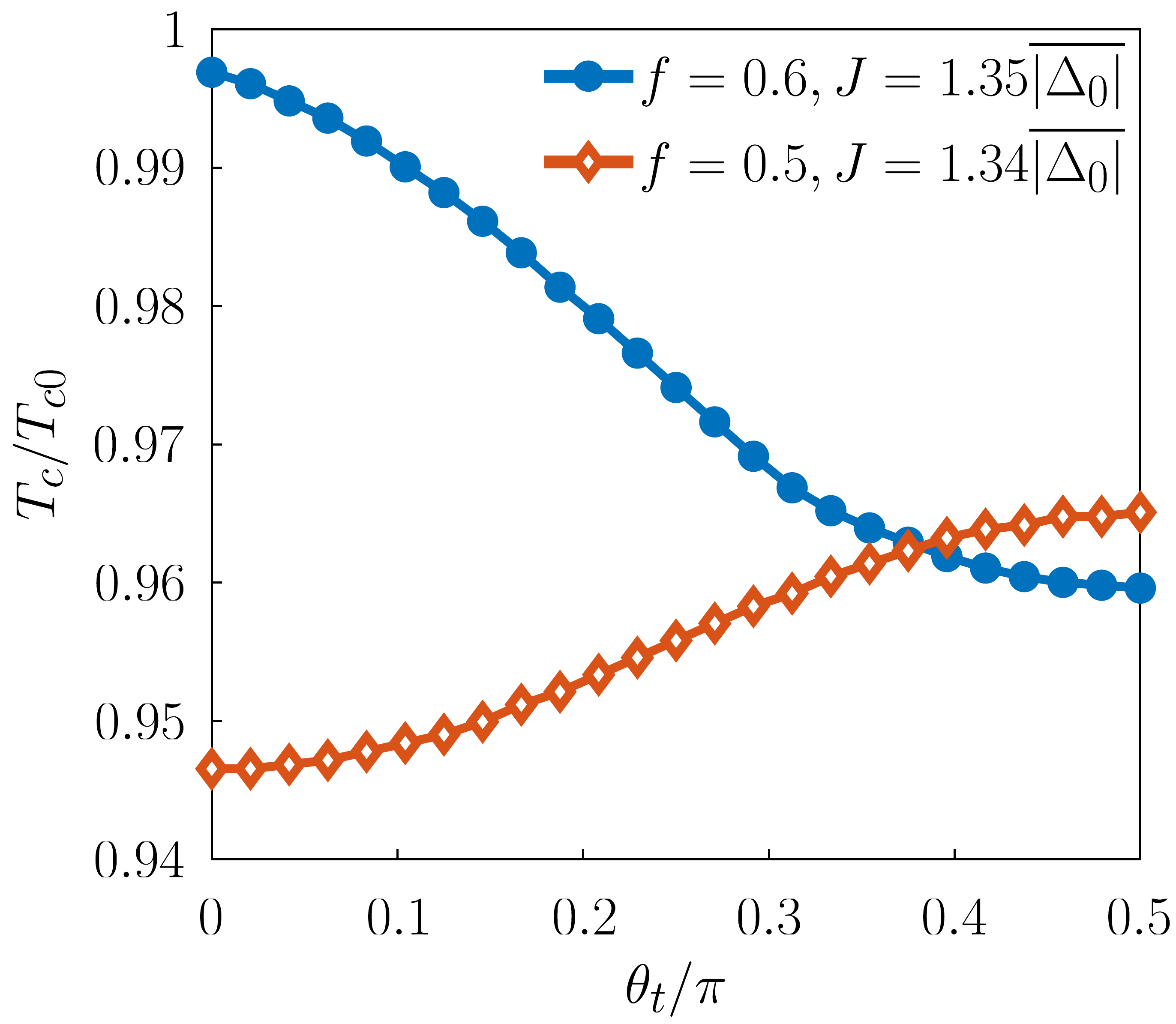}
		\caption{\lina{The critical temperature $T_c$ vs.~canting angle $\theta_t$ is plotted for a  2-D superconductor of size $302 \times 3$ interfaced with a canted-AF \add{at filling factors $f=0.5$ and $f=0.6$}. $T_{c0}$ and $\overline{|\Delta_0|}$ are the critical temperature and the average superconducting order parameter at zero-temperature, respectively, in the absence of the magnetic layer.} The detailed parameters employed for the numerical evaluation are specified in Appendix~\ref{parameter}.}
		\label{fig:2D_Tcplots}
	\end{center}
\end{figure*} 

\lina{In section~\ref{sec:Tc}}, we found that the \lina{critical} temperature vs.~canting angle curves \lina{of a 1-D superconductor} show opposite trends \lina{at} half-filling ($f=0.5$) and away from half-filling \lina{($f= 0.6$)}. 
\lina{In Fig.~\ref{fig:2D_Tcplots}, we show that $T_c$ behaves similarly for a 2-D superconductor. The} change in $T_c$ as we change the canting angle from the collinear AF \lina{case ($\theta_t=0$)} to the \lina{F} case ($\theta_t=\pi/2$) \lina{however} decreases in magnitude. \ak{This is because with an increasing S thickness, the effective spin-splitting induced by the AF decreases.}

}


\section{Numerical parameters} \label{parameter}
The parameters used for each of the figures are
\begin{itemize}
	\item \textbf{Fig.~\ref{fig:2}}: A 1-D superconductor with number of sites $N=302$, hopping parameter $t=10$, pair potential $U/t=1$, magnetic exchange interaction strength $J=0.018 \, t \, (=1.09\Delta_0)$, canting angle $\theta_t=\pi/4$, and filling factor $f=0.6$ $(\mu/\Delta_0=37)$, at temperature $T=0.1\, T_c=0.001 \, t/k_B$ is taken to calculate $s$-wave triplet correlations. Here, $2\Delta_0$ is the zero-temperature superconducting gap without the adjacent AF.
	
	\item \textbf{Fig.~\ref{fig:3}}:  A 1-D superconductor with $N=302$, $t=10$, and $U/t=1$ at $k_B T/t=0.001$ is taken for calculating correlations. For filling factor $f=0.5 \, (\mu/\Delta_0=0)$, $J=0.012 \, t=1.09\Delta_0$ is taken, and for $f=0.6 \, (\mu/\Delta_0=37)$, $J=0.018 \, t=1.09\Delta_0$, is taken. Here, $2\Delta_0$ represents the superconducting gap in the absence of the canted-AF at zero-temperature.
	
	\item \textbf{Fig.~\ref{fig:5}}: A 1-D superconductor with $N=302$, $t=10$, $U/t=1$ is used for the calculation of $T_c$ for filling factors $f=0.5 \, (\mu/\Delta_0=0)$ and $f=0.6 \, (\mu/\Delta_0=37)$. The critical temperature of the isolated superconductor $T_{c0}$ for $f=0.6$ is $0.0101 \, t/k_B$ and for $f=0.5$ is $0.0076 \, t/k_B$.
	
	\item \add{\textbf{Fig.~\ref{fig:4}}: A 2-D superconducting sheet with $N_x=102$, $N_y=14$, $t=10$, $J=0.1 \, t=4.3 \, \overline{|\Delta_0|}$, $\theta_t=\pi/4$, $U/t=1$, and $f=0.6 \, (\mu/\overline{|\Delta_0|}=18.5)$ at $k_B T/t=0.001$ is used to calculate the $s$-wave triplet correlations. Here, $\overline{|\Delta_0|}$ represents the average magnitude of the superconducting order parameter in absence of the canted-AF layer.}
	
	\item \add{\textbf{Fig.~\ref{fig:pycorrn}}: A 2-D superconductor with $N_x=202$, $N_y=14$, $t=10$, $J=0.1 \, t=4.3 \, \overline{|\Delta_0|}$, $\theta_t=\pi/4$, $U/t=1$, and $f=0.6 \, (\mu/\overline{|\Delta_0|}=18)$ at $k_B T/t=0.001$ is used to calculate the $p_y$-wave triplet correlations. Here, $\overline{|\Delta_0|}$ represents the average magnitude of the superconducting order parameter in absence of the AF.}
	
	\item \add{\textbf{Fig.~\ref{fig:corr_vs_jy}}: A 2-D superconductor with $N_x=102$, $N_y=30$, $t=10$, $J=0.1 \, t=1.7\, \overline{|\Delta_0|}$, $\theta_t=\pi/4$, $U/t=1$, and $f=0.5 \, (\mu=0)$ at $k_B T/t=0.001$ is used to calculate the $s$-wave triplet correlations, where $\overline{|\Delta_0|}$ represents the average magnitude of the superconducting order parameter in absence of the AF.}
	
	\item \add{\textbf{Fig.~\ref{fig:2D_Tcplots}}: A 2-D superconductor with $N_x=302$, $N_y=3$, $t=10$, and $U/t=1$ is used to plot $T_c$ vs.~$\theta_t$. For $f=0.5$, $J=0.047 \, t=1.34 \, \overline{|\Delta_0|}=2.4 \, T_{c0}$ is taken, and for $f=0.6$, $J=0.07 \, t=1.35 \, \overline{|\Delta_0|}=2.4 \, T_{c0}$ is taken, where $T_{c0}$ and $\overline{|\Delta_0|}$ are the critical temperature and the average magnitude of the superconducting order parameter at zero temperature of the same superconductor without the AF layer.}
\end{itemize}


\section{Canted-AF/S with rotated magnetic moments} \label{axischange}

\begin{figure*}[tb]
	\begin{center} 
		\includegraphics[width=\linewidth]{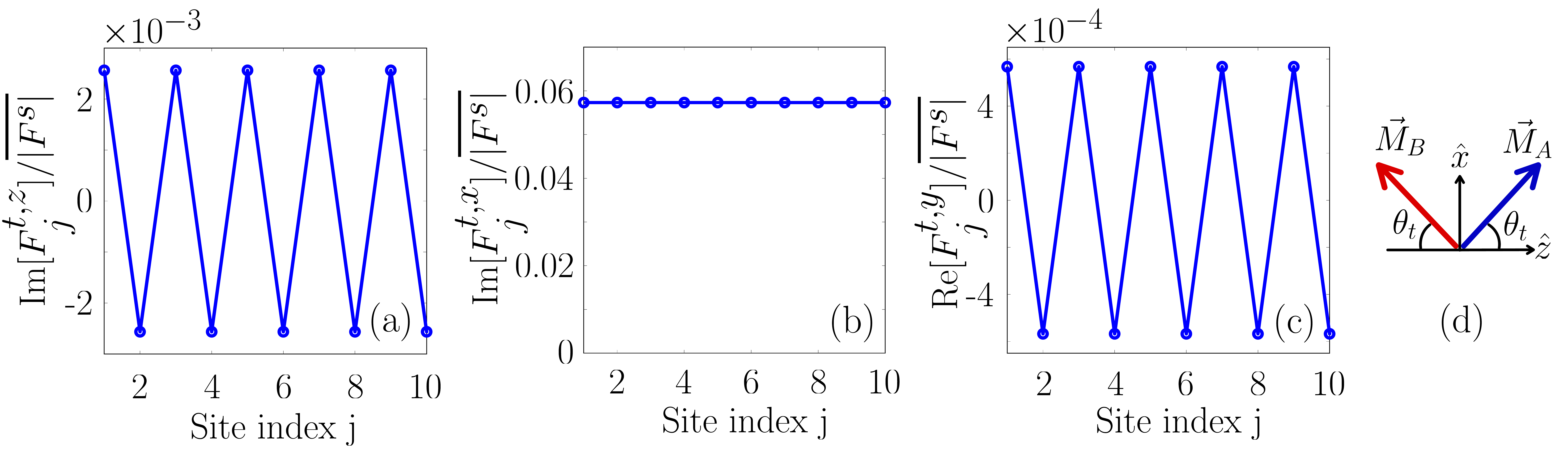}
		\caption{Spatial variation of normalized triplet correlations for 10 lattice sites considering $\theta_t=\pi/4$ and $\mu/\Delta_0=37$, corresponding to filling factor $f=0.6$. We show the imaginary part of the triplet correlations $F^{t,z}_j$ [panel (a)] and $F^{t,x}_j$ [panel (b)], and the real part of the triplet correlation $F^{t,y}_j$ [panel (c)]. The real part of the two former and the imaginary part of the latter are zero. All the correlations are normalized by the spatially averaged magnitude of the singlet correlation $\overline{ |F^s| }$. (d) Orientation of the magnetic moments of sublattices A and B of the AF in the rotated configuration.}
		\label{neelt}
	\end{center}
\end{figure*}

To compare the results of this article with Ref.~\cite{Bobkov2022}, one needs to rotate the magnetic moments of the Hamiltonian [Eq.~\eqref{eq1}] to $\vec{M}_j=\left[(-1)^{j+1}\cos \theta_t \, \hat{z} + \sin \theta_t \, \hat{x}\right]$ (see Fig.~\ref{neelt}(d)). The correlations for this rotated system is plotted in Fig.~\ref{neelt}. Here, $F^{t,z}$ is the N\'{e}el triplet correlation coming from the antiferromagnetic component of the canted-AF, $F^{t,x}$ comes from the ferromagnetic component, and $F^{t,y}$ comes from the noncollinearity in the canted-AF.

\section*{Acknowledgments}
SC and AK acknowledge financial support from the Spanish Ministry for Science and Innovation -- AEI Grant CEX2018-000805-M (through the ``Maria de Maeztu'' Programme for Units of Excellence in R\&D). LJK acknowledges support from the Research Council of Norway through its Centres of Excellence funding scheme, Project No. 262633 ‘QuSpin’. IVB acknowledges support from MIPT, Project FSMG-2023-0014.

\bibliography{Cant_AFS}

\end{document}